\documentstyle[12pt, epsf, amstex]{article}
\begin{document}

\pagestyle{plain} \setcounter{page}{1}
\newcounter{bean}
\baselineskip16pt

\begin{titlepage}
\begin{flushright}
SU-ITP 99/39 \\ KUL-TF-99/30 \\ hep-th/9908056
\end{flushright}
\vspace{7 mm}
\begin{center}
{\huge Magnetic fields, branes and \\ \vspace*{1mm} noncommutative
geometry} \vspace{5mm} {\huge }
\end{center}
\begin{center}
{\large D.~Bigatti$^\dagger$ and L.~Susskind$^\ddagger$ } \\
\vspace{13mm}
\vspace{7mm} {\large Abstract} \\
\end{center}
\vspace{3mm} We construct a simple physical model of a particle
moving on the infinite noncommutative 2-plane. The model consists
of a pair of opposite charges moving in a strong magnetic field.
In addition, the charges are connected by a spring. In the limit
of large magnetic field, the charges are frozen into the lowest
Landau levels. Interactions of such particles
include Moyal-bracket phases characteristic
of field theories on noncommutative space. The simple system
arises in the light cone quantization of open strings attached to
D-branes in antisymmetric tensor backgrounds. We use the model to
work out the general form of light cone vertices from string
splitting. We then consider the form of Feynman diagrams in
(uncompactified) noncommutative Yang Mills theories. We find that
for all planar diagrams the commutative and noncommutative
theories are exactly the same apart from trivial external line
factors. This means that the large $N$ theories are equivalent in
the 't Hooft limit. Non-planar diagrams are generally more
convergent than their commutative counterparts.

\noindent \vspace{1mm}
\begin{flushleft}
August 1999 \\ \vspace{1cm} $\dagger$ KU Leuven, Belgium \\ $\ddagger$
Stanford University, USA
\end{flushleft}
\end{titlepage}

\newpage
\renewcommand{\baselinestretch}{1.1}
\section{The model}
Gauge theories on noncommutative spaces \cite{1}\cite{2} are believed to
be relevant to the quantization of D-branes in background $B_{\mu
\nu}$ fields \cite{3}. The structure of such theories is similar
to that of ordinary gauge theory except that the usual product of
fields is replaced by a ``star product'' defined by
\begin{eqnarray} \displaystyle{
\phi * \chi = \phi(X) \exp\{-i \theta^{\mu \nu}{\partial \over
\partial X^{\mu}}
{\partial \over \partial Y^{\nu} }\}
\chi(Y)}
\end{eqnarray}
where $\theta^{\mu \nu}$ is an antisymmetric constant tensor. The
effect of such a modification is reflected in the momentum space
vertices of the theory by factors of the form
\begin{eqnarray}
\label{cippa2} \displaystyle{ \exp[i \theta^{\mu \nu} p_\mu q_\nu ]
} \equiv e^{i{p\wedge q}}
\end{eqnarray}
The purpose of this paper is to show how these factors arise in an
elementary way. We will begin by describing a simple quantum
mechanical system which is fundamental to our construction. We
then consider string theory in the presence of a D3-brane and a
constant large $B_{\mu \nu}$ field. In the light cone frame the
first quantized string is described by our elementary model. We
use the model to compute the string splitting vertex and show how
the factors in eq.~(2) emerge. We then turn to the structure of the
perturbation series for the non-commutative theory in infinite
flat space. We find that  planar diagrams with any number of
loops are identical to their commutative counterparts apart from
trivial external line phase factors.

Compactification, which can lead to entirely new features, is not
studied in this paper.

\subsection{The model at classical level}
Consider a pair of unit charges of opposite sign in a magnetic
field $B$ in the regime where the Coulomb and the radiation terms
are negligible. 
The coordinates of the charges are $\vec{x_1}$ and
$\vec{x_2}$ or in component form $x_1^i$ and $x_2^i$. The
Lagrangian is
\begin{eqnarray} \label{1.1} \displaystyle{
{\cal L} = \frac{m}{2} \left( (\dot{x}_1)^2 + (\dot{x}_2)^2
\right) \:+\: \frac{B}{2} \epsilon_{ij} \left( \dot{x}_1^i x_1^j -
\dot{x}_2^i x_2^j \right) \: -\: \frac{K}{2} (x_1-x_2)^2 }
\end{eqnarray} where the first term is the kinetic energy of the
charges, the second term is their interaction with the magnetic
field and the last term is an harmonic potential between the
charges.
\par In what follows we will be interested in the limit in which
the first term can be ignored. This is typically the case if $B$
is so large that the available energy is insufficient to excite
higher Landau levels \cite{girvin}. 
Thus we will focus on the simplified Lagrangian
\begin{eqnarray} \displaystyle{
{\cal L} = \frac{B}{2} \epsilon_{ij} \left( \dot{x}_1^i x_1^j -
\dot{x}_2^i x_2^j \right) \: -\: \frac{K}{2} (x_1-x_2)^2 }
\end{eqnarray} Let us first discuss the classical system. The
canonical momenta are given by
\begin{eqnarray} \displaystyle{
\begin{array}{l}
p_i^1 = \displaystyle{ \frac{\partial {\cal L}}{\partial
\dot{x}_1^i} } = B \epsilon_{ij} x_1^j \\ p_i^2 = - B
\epsilon_{ij} x_2^j
\end{array}
} \end{eqnarray} Let us define center of mass and relative
coordinates $X$, $\Delta$:
\begin{eqnarray} \displaystyle{
\begin{array}{l}
\vec{X} = (\vec{x_1} + \vec{x_2}) /2 \\ \vec{\Delta} = (\vec{x_1}
- \vec{x_2}) /2
\end{array}
} \end{eqnarray} The Lagrangian is
\begin{eqnarray} \label{5} \displaystyle{
{\cal L} = m((\dot{X})^2+(\dot{\Delta})^2) + 2 B \epsilon_{ij}
\dot{X}^i \Delta^j - 2 K (\Delta)^2 } \end{eqnarray} Dropping the
kinetic terms gives
\begin{eqnarray} \displaystyle{
{\cal L} = 2 B \epsilon_{ij} \dot{X}^i \Delta^j - 2 K (\Delta)^2 }
\end{eqnarray} The momentum conjugate to $X$ is
\begin{eqnarray} \label{1.6} \displaystyle{
\frac{\partial {\cal L}}{\partial \dot{X}^i} = 2 B \epsilon_{ij}
\Delta^j = P_i } \end{eqnarray} This is the center of mass
momentum.
\par Finally, the Hamiltonian is
\begin{eqnarray} \label{7} \displaystyle{
{\cal H} = 2 K (\Delta)^2 = 2 K \left( \frac{P}{2B} \right)^2 =
\frac{K}{2B^2} P^2 } \end{eqnarray} This is the hamiltonian of a
nonrelativistic particle with mass
\begin{eqnarray} \displaystyle{
M= \frac{B^2}{K} } \end{eqnarray}

Evidently the composite system of opposite charges
moves like a galileian particle of mass $M$. What is unusual is
that the spatial extension $\Delta$ of the system is related to
its momentum so that the size grows linearly with $P$ according to
eq.~(\ref{1.6}). How does this growth with momentum effect the interactions
of the composite? Let's suppose
charge 1 interacts  locally with an ``impurity'' centered at the origin.
The interaction  has the form
\begin{eqnarray} \label{9} \displaystyle{
V(\vec{x}_1) = \lambda \: \delta (\vec{x}_1) } \end{eqnarray} In
terms of $X$ and $\Delta$ this becomes
\begin{eqnarray} \label{10} \displaystyle{
V = \lambda \: \delta (X+\Delta) = \lambda \: \delta(X^i -
\frac{1}{2B} \epsilon^{ij} P_j ) } \end{eqnarray} Note that the
interaction in terms of the center of mass coordinate is nonlocal
in a particular way. The interaction point is shifted by a
momentum dependent amount. This is the origin of the peculiar
momentum dependent phases that appear in interaction vertices on
the noncommutative plane. More generally, if particle 1 sees a
potential $V(x_1)$ the interaction becomes
\begin{eqnarray} \label{1.7} \displaystyle{
V\left(X -\frac{\epsilon P}{2B} \right) } \end{eqnarray}

\subsection{Quantum level}
The main problem in quantizing the system is to correctly define
expressions like (\ref{1.7}) which in general have factor ordering
and other quantum ambiguities. In order to define them, let us
assume that $V$ can be expressed as a Fourier transform
\begin{eqnarray} \label{c14}  \displaystyle{
V(x) = \int dq  \: \tilde{V}(q) e^{iqx} } \end{eqnarray} We can then
formally write
\begin{eqnarray} \displaystyle{
V(X-\frac{\epsilon P}{2B}) = \int dq \: \tilde{V}(q) e^{i q (X-
\frac{\epsilon P}{2B}) } }  \end{eqnarray} The factor ordering is
not ambiguous because
\begin{eqnarray} \displaystyle{
[q_i X^i, q_l \epsilon^{lj} P_j] = q_i q_l \epsilon^{lj} [X^i P_j]
= 0 } \end{eqnarray} Consider the matrix element
\begin{eqnarray} \displaystyle{
\langle k |\exp[iq(X-\frac{\epsilon p}{2B})] | l \rangle }
\end{eqnarray} where $\langle k|$ and $|l \rangle$ are momentum
eigenvectors. Using eq.(17) we can write this as
\begin{eqnarray} \displaystyle{
\langle k | \exp[iqX] \: \exp[-i \frac{q \epsilon P}{2B}] | l \rangle
} \end{eqnarray} Since $|l\rangle$ is an eigenvector of $P$ this
becomes
\begin{eqnarray} \displaystyle{
\langle k |\exp[iqX] |l \rangle \exp[-i \frac{q \epsilon  l}{2B}]=
\delta(k-q-l) \exp[-i {q \epsilon l}/{2B}] } \end{eqnarray} The
phase factor is the usual Moyal bracket phase that is ubiquitous
in noncommutative geometry.

\section{String theory in magnetic fields}
Let us consider bosonic string theory in the presence of a
D3-brane. The coordinates of the brane are $x^0$, $x^1$, $x^2$,
$x^3$. The remaining coordinates will play no role. We will also
assume a background antisymmetric tensor field $B_{\mu \nu}$ in
the 1,2 direction. We will study the open string sector with
string ends attached to the D3-brane in the light cone frame.
\par Define
\begin{eqnarray} \displaystyle{
x^{\pm}= x^0 \pm x^3 } \end{eqnarray} and make the usual light
cone choice of world sheet time
\begin{eqnarray} \displaystyle{
\tau = x^+ } \end{eqnarray}
\par The string action is
\begin{eqnarray} \displaystyle{
{\cal L} = \frac{1}{2} \int_{-L}^L  \: d\tau \: d\sigma \: \left[
\left( \frac{\partial x^i}{\partial \tau} \right)^2 - \left(
\frac{\partial x^i}{\partial \sigma} \right)^2 + B_{ij}
\left( \frac{\partial x^i}{\partial \tau} \right) \left(
\frac{\partial x^j}{\partial \sigma} \right) \right] }
\end{eqnarray} We have numerically fixed $\alpha'$ and the parameter $L$ can
be identified with $P_-$, the momentum conjugate to $x_-$.
\par In what follows we will be interested in the limit $B \longrightarrow
\infty$. Let us make the following rescalings
\begin{eqnarray} \displaystyle{
\left\{
\begin{array}{l}
\displaystyle{x^i= \frac{y^i}{\sqrt{B}}} \\ \tau = tB
\end{array}
\right. } \end{eqnarray} Then

\begin{eqnarray} \displaystyle{
{\cal L} = \frac{1}{2} \int_{-L}^L  \: d\tau \: d\sigma \: \left[
\frac{1}{B^2} \left( \frac{\partial y}{\partial t} \right)^2 -
\left( \frac{\partial y}{\partial \sigma} \right)^2 +
\epsilon_{ij}  \left( \frac{\partial y^i}{\partial t} \right)
\left( \frac{\partial y^j}{\partial \sigma} \right) \right] }
\end{eqnarray}
\par Now for $B \longrightarrow \infty$ we can drop the first term.
Furthermore by an integration by parts and up to a total time derivative
the last term can be written
\begin{eqnarray} \label{c23} \displaystyle{ \left.
\epsilon_{ij} \frac{\partial y_i}{\partial t} y_j \right|^L_{-L} }
\end{eqnarray} Thus
\begin{eqnarray} \displaystyle{
{\cal L}= \frac{1}{2} \int d\sigma \; d\tau \left( \frac{\partial
y}{\partial \sigma} \right)^2 + \left. \epsilon_{ij} \dot{y_i} y_j
\right|_{-L}^L } \end{eqnarray} Since for $\sigma \neq \pm L$ the
time derivatives of $y$ do not appear in $S$ we may trivially
integrate them out. The solution of the classical equation of
motion is
\begin{eqnarray} \displaystyle{
y(\sigma) = y + \frac{ \Delta \sigma}{L} } \end{eqnarray}
with $\Delta$ and $y$ independent of $\sigma$. The resulting action is
\begin{eqnarray} \displaystyle{
{\cal L}=  \left[ -\frac{2 \Delta^2}{L} + \dot{y} \epsilon \Delta
\right] } \end{eqnarray} Evidently, the action is of the same form
as the model in section 1 with $B$ and $K$ rescaled.

\section{The interaction vertex}
Interactions in light cone string theory are represented by string
splitting and joining. Consider two incoming strings with momenta
$p_1$, $p_2$ and center of mass positions $y_1$, $y_2$. If their
endpoints coincide they can join to form a third string with
momentum $-p_3$. The constraints on the endpoints are summarized
by the overlap $\delta$ function
\begin{eqnarray}  \displaystyle{
{\Large \nu} = \delta((y_1 -  \Delta_1)-(y_2+ \Delta_2))  \: 
 \delta((y_2- \Delta_2) - (y_3 +  \Delta_3))  \: 
 \delta((y_3 -\Delta_3) - (y_1 +\Delta_1))
 } \end{eqnarray}
\par From eq.~(29) we see that the center of mass momentum is
related to $\Delta$ by
\begin{eqnarray} \displaystyle{
P=\epsilon \Delta } \end{eqnarray} Inserting this in eq.~(30) 
gives the vertex
\begin{eqnarray}
\displaystyle{
{\Large \nu} = \delta(y_1-y_2+ (\epsilon p_1 +\epsilon
p_2)) \:  \delta(y_2-y_3+ (\epsilon p_2 +\epsilon p_3)) \: 
\delta(y_3-y_1+ (\epsilon p_3 +\epsilon p_1)) }
\end{eqnarray}
\par To get the vertex in momentum space multiply by $\displaystyle{
e^{i(p_1 y_1 + p_2 y_2 + p_3 y_3)} }$ and integrate over $y$. This
yields
\begin{eqnarray} \label{c30}  \displaystyle{
{\Large \nu}= e^{i (p_1 \epsilon p_2) } \delta(p_1 +p_2
+p_3) } \end{eqnarray} This is the usual form of the vertex in
noncommutative field theory. We have scaled the  ``transverse''
coordinates $x^1$, $x^2$ (but not $x^0$, $x^3$) and momenta so
that the $B$ field does not appear in the vertex. If we go back to
the original units the phases in (\ref{c30}) will be proportional
to $1/B$.
\par Evidently a quantum of noncommutative Yang Mills theory may be thought
of as a
straight string connecting two opposite charges. The separation
vector $\Delta$ is perpendicular to the direction of motion $P$.
\par Now consider the geometry of the 3-body vertex. The string endpoints
$u$, $v$, $w$ define a triangle with sides
\begin{eqnarray} \displaystyle{
\begin{array}{c}
\Delta_1 = (u-v) \\ \Delta_2= (v-w) \\ \Delta_3=(w-u)
\end{array}
} \end{eqnarray} and the three momenta are perpendicular to the
corresponding $\Delta$. It is straightforward to see that the
phase
\begin{eqnarray} \displaystyle{
\epsilon_{ij} p_i q_j /B }\equiv p\wedge q \end{eqnarray} is just
the area of the triangle times $B$. In other words, it is the
magnetic flux through the triangle. Note that it can be of either
sign.
\par More generally, we may consider a Feynman tree diagram constructed
from such vertices. For example consider figure (1a). The overall
phase is the total flux through the triangles A, B and C. In fact we
can simplify this by shrinking the internal propagators to get
figure (1b). Thus the phase is the flux through a polygon formed
from the $\Delta$'s of the external lines. The phase depends only
on the momenta of the external lines and their cyclic order.

\section{Structure of Perturbation Theory}
In this section we will consider the effects of the Moyal phases
on the structure of Feynman amplitudes in noncommutative Yang
Mills theory. Let us first review the diagram rules for ordinary
Yang Mills theory in 't Hooft double-line representation.
\par The gauge propagator can be represented as a double line as if
the gauge boson were a quark-antiquark pair as in figure (2). Each
gluon is equipped with a pair of gauge indices $i,j$, a momentum
$p$ and a polarization $\varepsilon$ satisfying $\varepsilon \cdot p=
\varepsilon^\mu p_\mu =0$. 
\par The vertex describing 3-gauge boson interaction is shown in
figure (3). In addition to Kronecker $\delta$ for the gauge indices
and momentum $\delta$ functions the vertex contains the factor
\begin{eqnarray} \displaystyle{
(\varepsilon_1 \cdot p_3 + \varepsilon_3 \cdot p_2 + \varepsilon_2 \cdot
p_1 ) } \end{eqnarray} The factor is antisymmetric under
interchange of any pair and so it must be accompanied by an
antisymmetric function of the gauge indices. For a purely abelian
theory the vertex vanishes when symmetrized.
\par Now we add the new factor coming from the Moyal bracket. This
factor is
\begin{eqnarray} \displaystyle{
e^{i p_1 \wedge p_2} = e^{i p_2 \wedge p_3} = e^{i p_3 \wedge p_1}
} \end{eqnarray} where $p_a \wedge p_b$ indicates an antisymmetric
product
\begin{eqnarray} \displaystyle{
\begin{array}{c}
p \wedge q = p_\mu q_\nu \theta^{\mu \nu} \\ \theta^{\mu \nu} = -
\theta^{\nu \mu}
\end{array}
} \end{eqnarray} Because these factors are not symmetric under
interchange of particles, the vertex no longer vanishes when Bose
symmetrized  even for the Abelian theory.
\par The phase factors satisfy an important identity. Let us consider the
phase factors that accompany a given diagram. In fact from now on
a diagram will indicate {\bf only the phase factor} from the
product of vertices. Now consider the diagram in figure (4a). It is
given by
\begin{eqnarray} \displaystyle{
e^{i(p_1 \wedge p_2)} e^{i (p_1 +p_2) \wedge p_3} = e^{i (p_1
\wedge p_2 + p_2 \wedge p_3 + p_1 \wedge p_3)} } \end{eqnarray} On
the other hand the dual diagram figure (4b) is given by $e^{i(p_1
\wedge (p_2 + p_3) + p_2 \wedge p_3)}$. It is identical to the
previous diagram. Thus the Moyal phases satisfy  old fashioned
``channel duality''. This conclusion is also obvious from the
``flux through polygon'' construction of the previous section.
\par In what follows, a ``duality move'' will refer to a replacement of a
diagram such as in figure (4a) by the dual diagram in figure (4b).
\par Now consider any planar diagram with $L$ loops. By a series of
``duality moves'' it can be brought to the form indicated in figure (5)
consisting of a tree with $L$ simple one-loop tadpoles.
\par Let us consider the tadpole, figure (6).
The phase factor is just $e^{i q \wedge q}=1$. Thus the loop
contributes nothing to the phase and the net effect of the Moyal
factors is exactly that of the tree diagram. In fact all trees
contributing to a given topology have the same phase, which is a
function only of the external momentum. The result is that for
planar diagrams the Moyal phases do not affect the Feynman
integrations at all. In particular the planar diagrams have
exactly the same divergences as in the commutative theory.
Evidently in the large N limit noncommutative Yang Mills =
ordinary Yang Mills.
\par On the other hand, divergences that occur in nonplanar diagrams
can be regulated by the phase factors. For example consider the
nonplanar diagram in figure (7). The Moyal phase for the diagram is
\begin{eqnarray} \displaystyle{
e^{i p \wedge q} e^{i p \wedge q} = e^{2 i p \wedge q} }
\end{eqnarray} and does not cancel. It is not difficult to see that
such oscillating phases will
regulate divergent diagrams and make them finite, unless the
diagram contains divergent planar subdiagrams. Thus it seems that
the leading  high momentum behavior of the theory is controlled by the
planar diagrams. Among other things it means that in this region
the $1/N$ corrections to the 't Hooft limit must vanish.

An interesting question arises if we study the theory on a torus
of finite size \cite{dani}. For an ordinary local field theory high momentum
behavior basically 
corresponds to small distance behavior. For this reason we expect
the high momentum behavior on a torus to be identical to that in
infinite space once the momentum becomes much larger than the
inverse size of the torus. However in the noncommutative case the
story is more interesting. We have seen that high momentum in the
1,2 plane is associated with $large$ distances in the
perpendicular direction. Most likely this means that the finite
torus generically behaves very differently at high momentum than
the infinite plane.

Indeed there is an
exception to the rule that nonplanar diagrams are finite.
If   a line with a nonplanar self energy
insertion such as in figure (7) happens to have vanishing momentum
in the 1,2 plane then according to eq.~(40) its phase will vanish.
Thus for a set of measure zero, the nonplanar self energy
diagram can diverge. This presumably leads to no divergences in infinite space
when the line in question is integrated over. The situation could be
different for compact noncommutative geometries since integrals
over momenta are replaced by sums \cite{4}.

\par The fact that the large N limit is essentially the same
for noncommutative and ordinary Yang Mills theories implies that
in the AdS/CFT correspondence the introduction of noncommutative
geometry does not change the thermodynamics of the theory \cite{5}. It may
also be connected to the fact that in the matrix theory
construction of Connes-Douglas-Schwartz and Douglas-Hull, the
large N limit effectively decompactifies $X^{11}$ and should
therefore eliminate dependence on the 3-form potential. However
the argument is not straightforward since in matrix theory we are
not usually in the 't Hooft limit.

\section*{Acknowledgements}
L.~S.~would like to thank Steve Shenker for discussion.

\newpage 

\begin{figure}[h] 
\epsfxsize=150mm
\epsfbox{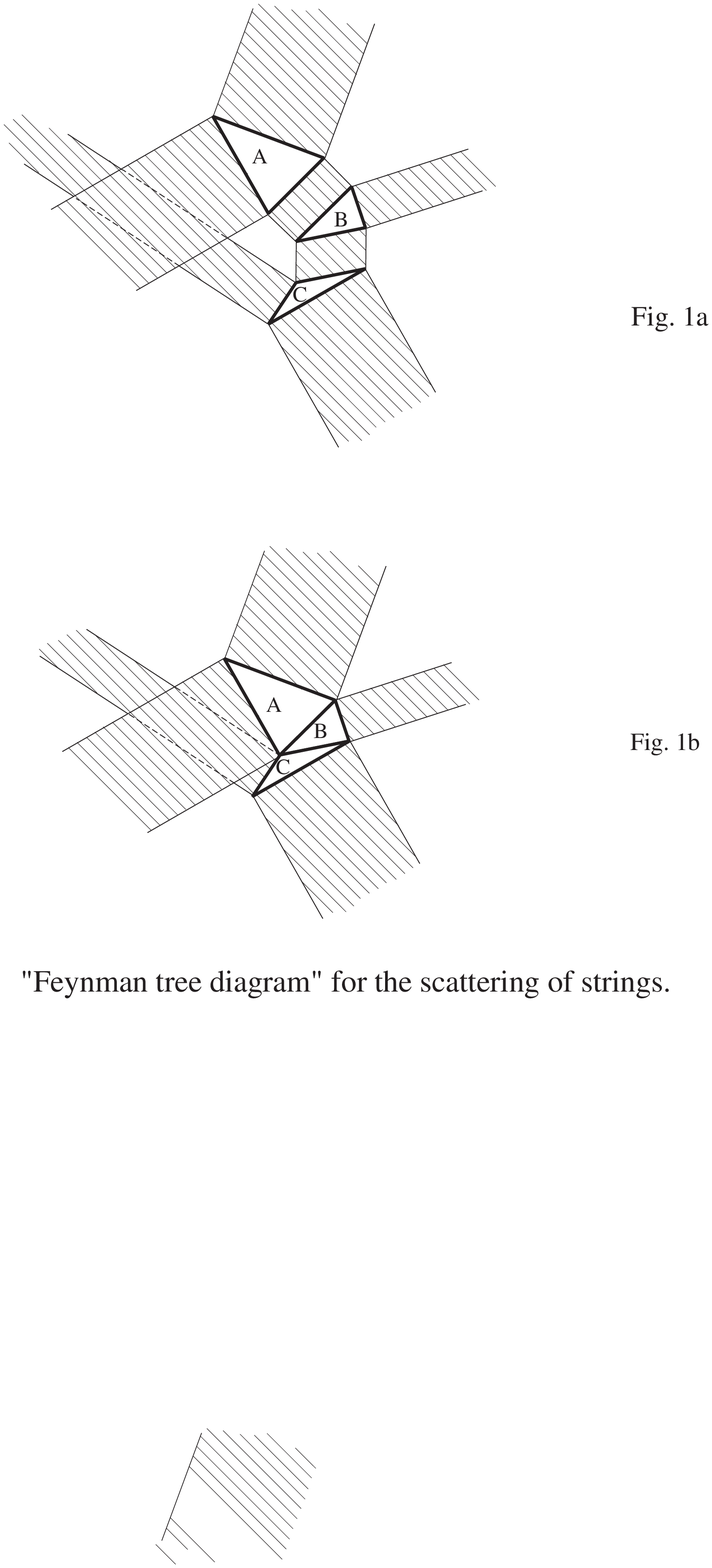}
\end{figure}

\begin{figure}[h] 
\epsfxsize=150mm
\epsfbox{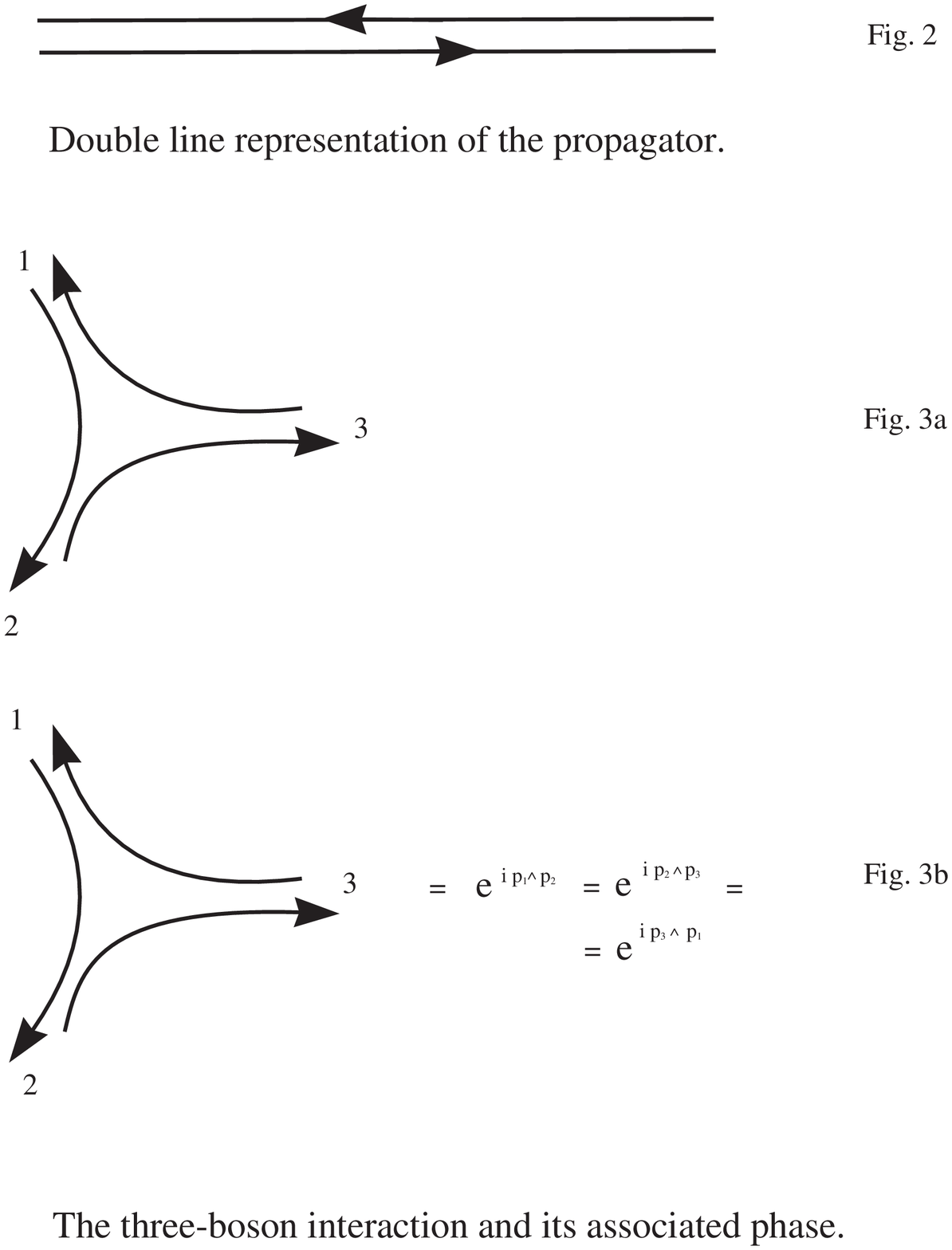}
\end{figure}

\begin{figure}[h]
\epsfxsize=150mm
\epsfbox{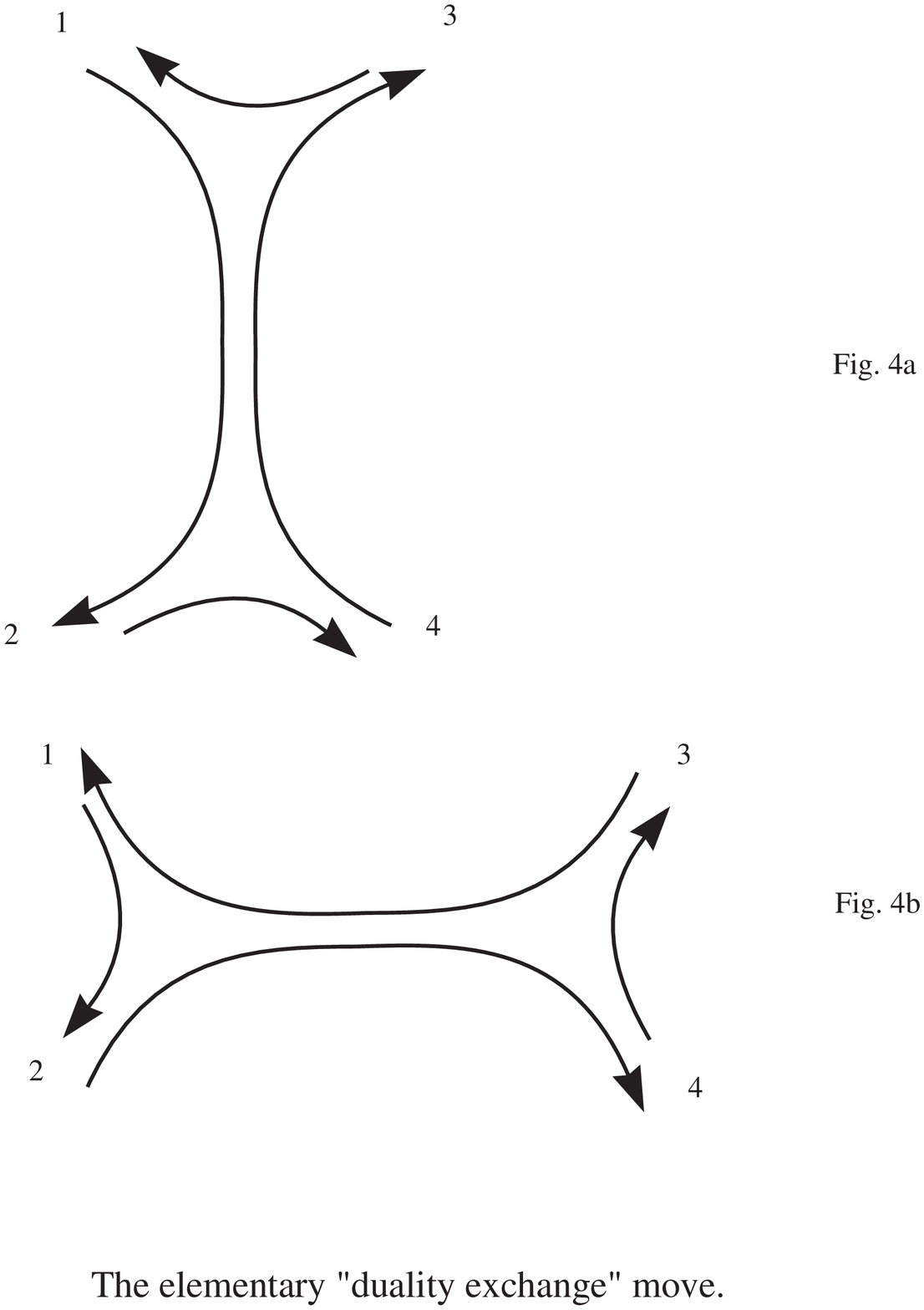}
\end{figure} 

\begin{figure}[h]
\epsfxsize=150mm
\epsfbox{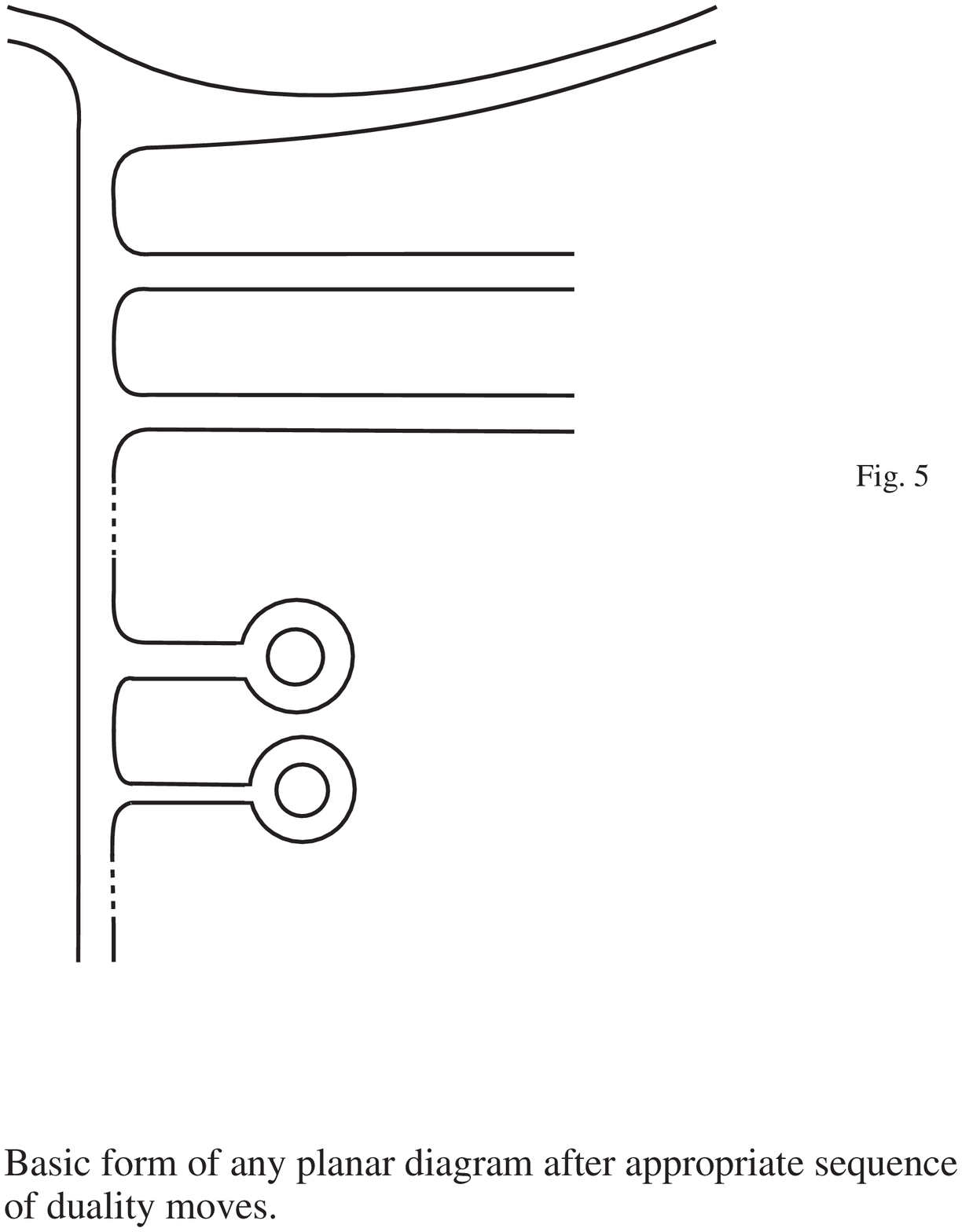}
\end{figure}

\begin{figure}[h]
\epsfxsize=150mm
\epsfbox{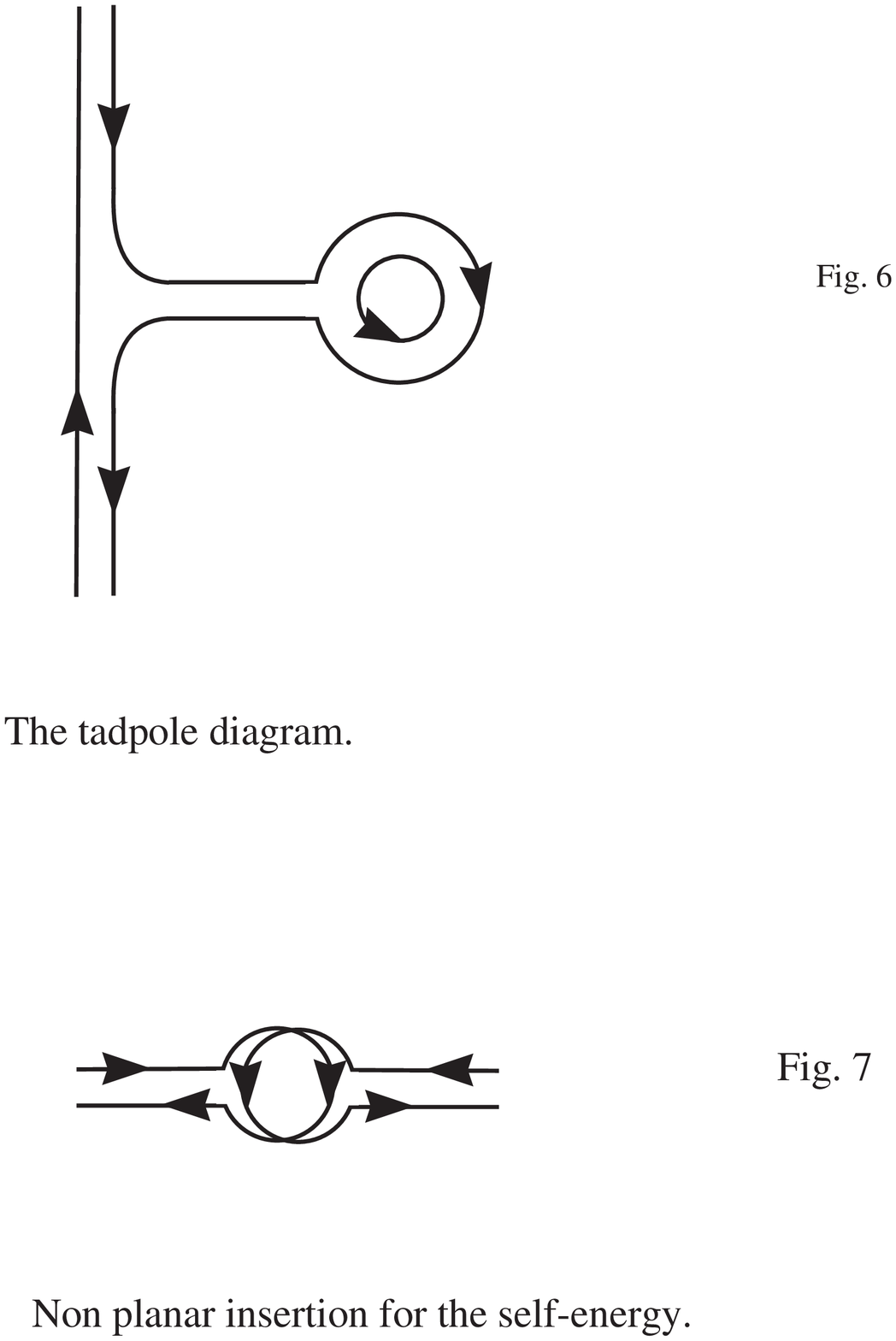}
\end{figure}

\end{document}